\documentclass[reprint,amsmath,amssymb,aps]{revtex4-2}    
\usepackage{graphicx}
\usepackage{xcolor}

\begin{document}

\title{Emergent noise-aided logic through synchronization}

\author{Manaoj Aravind}
\affiliation{Department of Physics, Indian Institute of Technology Bombay, Powai, Mumbai 400 076, India}

\author{Sudeshna Sinha}
\affiliation{Indian Institute of Science Education and Research Mohali, Knowledge City, SAS Nagar, Sector 81, Manauli, Punjab, PO 140 306, India}

\author{P. Parmananda}
\affiliation{Department of Physics, Indian Institute of Technology Bombay, Powai, Mumbai 400 076, India}

\date{\today}

\begin{abstract}
In this article, we present a dynamical scheme to obtain a reconfigurable noise-aided logic gate, that yields all six fundamental 2-input logic operations, including the XOR operation. The setup consists of two coupled bistable subsystems that are each driven by one subthreshold logic input signal, in the presence of a noise floor. The synchronization state of their outputs, robustly maps to 2-input logic operations of the driving signals, in an optimal window of noise and coupling strengths. Thus the interplay of noise, nonlinearity and coupling, leads to the emergence of logic operations embedded within the collective state of the coupled system. This idea is manifested using both numerical simulations and proof-of-principle circuit experiments. The regions in parameter space that yield reliable logic operations were characterized through a stringent measure of reliability, using both numerical and experimental data.
\end{abstract}

\maketitle

\section{Introduction}

Exploiting the richness in the behaviour of nonlinear dynamical systems for computational tasks has attracted extensive research interest. Various theoretical schemes have been proposed to realize reconfigurable devices using the rich patterns in chaotic systems and varied experimental implementations of these schemes have been been realized\cite{sinha1998dynamics,intro,sync}. 

An interesting recent line of enquiry is the effect of noise in such schemes. This approach is crucial as this leads to the possibility of noise-aided computational devices that can utilize noise to facilitate computing. Logical Stochastic Resonance (LSR)\cite{murali2009reliable,sinha2009exploiting,bulsara2010logical,lsr_noisefree,lsr_colorednoise,lsr_colorednoise2,memory,srlatch,forcing,das,fractional,weiner}, is one such scheme in which a bistable system driven by a subthreshold stream of inputs can generate responses consistent with a desired logical operation for an optimal window of noise. This again has been implemented in a wide variety of experimental systems\cite{apl,nano,jaimes,sgn_expt} that range from synthetic gene networks\cite{ando2011synthetic,sgn_dari,sgn_timedelay,sgn_adaptive,sgn_levy,sgn_wang}, optical systems\cite{lsr_optics1,lsr_optics2} to coulomb coupled quantum dots\cite{pfeffer2015logical}. More recent efforts have focused on using chaotic attractor hopping \cite{murali2018chaotic,murali2021harnessing} and strange non-chaos \cite{aravindh2018strange,sathish2020route} to yield logic gates. 

The effect of coupling on such noisy bistable systems has garnered new interest \cite{aravind2020synchronized,aravind2021competitive}. Coupling induced LSR \cite{aravind2018coupling} demonstrated that logic operations maybe obtained from the state variable of coupled bistable systems in an optimal window of noise. 

In this work, we show a novel possibility that the logical output can be embedded in collective state of the coupled system, rather than in the state variable of one of the subsystems. This change leads to a rich construct, which can yield all 2-input logic operations for an optimal window of noise strength and coupling strength. 

\section{Scheme}

\begin{figure}
\centering
\includegraphics[width=0.9\linewidth]{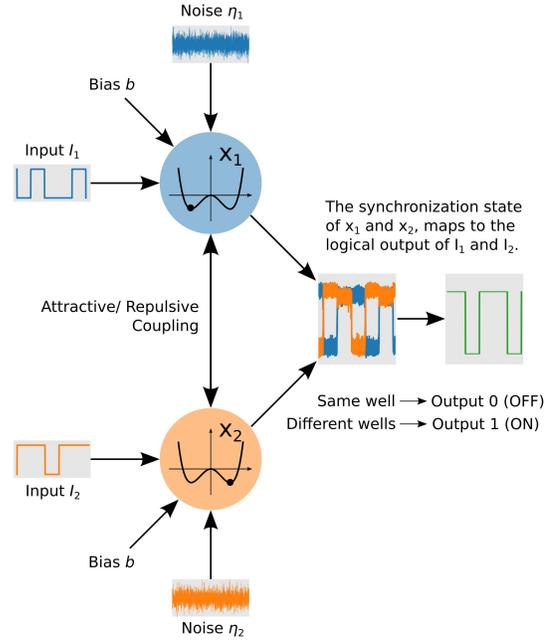}
\caption{A schematic representation of the concept.
}
\label{fig:schematic}
\end{figure}

Consider coupled bistable systems with two kinds of coupling schemes defined as follows (refer Fig.~\ref{fig:schematic}), 

\begin{eqnarray}
\nonumber
\dot{x_1} = F (x_1) +  c(x_2-x_1) + I_1(t) + b +  D \ \eta_1 (t) \\ 
\dot{x_2} = F (x_2) +  c(x_1-x_2) - I_2(t) - b + D \ \eta_2 (t)
\label{eq:att_model}
\end{eqnarray}
and
\begin{eqnarray}
\nonumber
\dot{x_1} = F (x_1) +  c(-x_2-x_1) + I_1(t) + b +  D \ \eta_1 (t) \\ 
\dot{x_2} = F (x_2) +  c(-x_1-x_2) + I_2(t) + b + D \ \eta_2 (t)
\label{eq:rep_model}
\end{eqnarray}

Here, F can be any nonlinear function that yields a bistable potential. The terms $\eta_1$ and $\eta_2$ are two uncorrelated, zero mean, univariate Gaussian noises of noise strength $D$. The drive signals $I_1$ and $I_2$ are two low amplitude (subthreshold) input streams that encode the two binary inputs to the system. A constant asymmetrizing bias $b$ acts as the tether that morphs the bistable potential, leading to reconfigurability in our scheme. In the above equations, two kinds of coupling terms have been used. In Eq.~\ref{eq:att_model}, the coupling between the two subsystems is bidirectional and attractive in nature with coupling strength $c$. This typical diffusive kind of coupling interaction \cite{stankovski2017coupling} tends to synchronize the subsystems i.e. in the context of bistable systems, the two subsystems are pulled to the same potential well. In Eq.~\ref{eq:rep_model}, the coupling interaction is bidirectional and repulsive in nature. This form of coupling repels the states of the subsystems, thus tending to anti-synchronize the subsystems \cite{aravind2021competitive,hens2013oscillation,dixit2020static}, i.e. for bistable systems, the subsystems are pushed to different potential wells. Thus, Equation \ref{eq:att_model} describes \textit{attractive coupling} and Equation \ref{eq:rep_model} describes \textit{repulsive coupling}. In conjunction with the bias $b$, we demonstrate that changing between these two coupling forms offers another degree of control, that allows us to obtain all six fundamental logic operations (c.f. Table~\ref{tab:reconf}). 

\begin{table*}
\caption{Relationship between the two inputs and the output of the fundamental OR, AND, NOR, NAND, XOR and XNOR operations, for the four distinct possible input sets $(0,0)$, $(0,1)$, $(1,0)$ and $(1,1)$.}
\label{tab:gates} 
\begin{center}
\begin{tabular}{p{4.0cm}p{1.5cm}p{1.5cm}p{1.5cm}p{1.5cm}p{1.5cm}p{1.5cm}}
\hline

Input Set ($I_1$,$I_2$) & OR & AND & NOR & NAND & XOR & XNOR \\
\noalign{\smallskip}\hline\noalign{\smallskip}
(0,0) & 0 & 0 & 1 & 1 & 0 & 1\\ 
(0,1)/(1,0) & 1 & 0 & 0 & 1 & 1 & 0\\ 
(1,1) & 1 & 1 & 0 & 0 & 0 & 1\\
\hline
\end{tabular}
\end{center}
\end{table*}

The signals $I_1$ and $I_2$ encode the stream of binary inputs to be processed by the logic gate, where without loss of generality $I_1$(or $I_2$) $ < 0$ corresponds to a $0$ (OFF state) and $I_1$(or $I_2$) $ > 0$ corresponds to a $1$ (ON state). It is important to note that, these input signals are subthreshold i.e., they cannot cause a transition between the potential wells on their own and the transitions are actively facilitated by the noise floor, reminiscent of logical stochastic resonance (LSR). The crucial differentiating factor of this scheme from all previous attempts to achieve noise-aided logic operations is that the output from the system is embedded in the collective state of the coupled system. Specifically, \textit{the synchronization state of the two subsystems embeds outputs corresponding to logical operations on the input streams}. As a convention, the synchronized state (both systems in same potential well) is taken to encode $0$ and the anti-synchronized state (both systems in different potential wells) is taken to encode $1$. The binary input output relations represented by the six fundamental 2-input logic gates are detailed in Table~\ref{tab:gates}. With these conventions in place, we now show that this scheme is capable of producing all the six fundamental logic operations in a robust, reliable manner over a large region of parameter space. 

\section{Implementation}

\begin{figure*}
\centering
\includegraphics[width=0.49\linewidth]{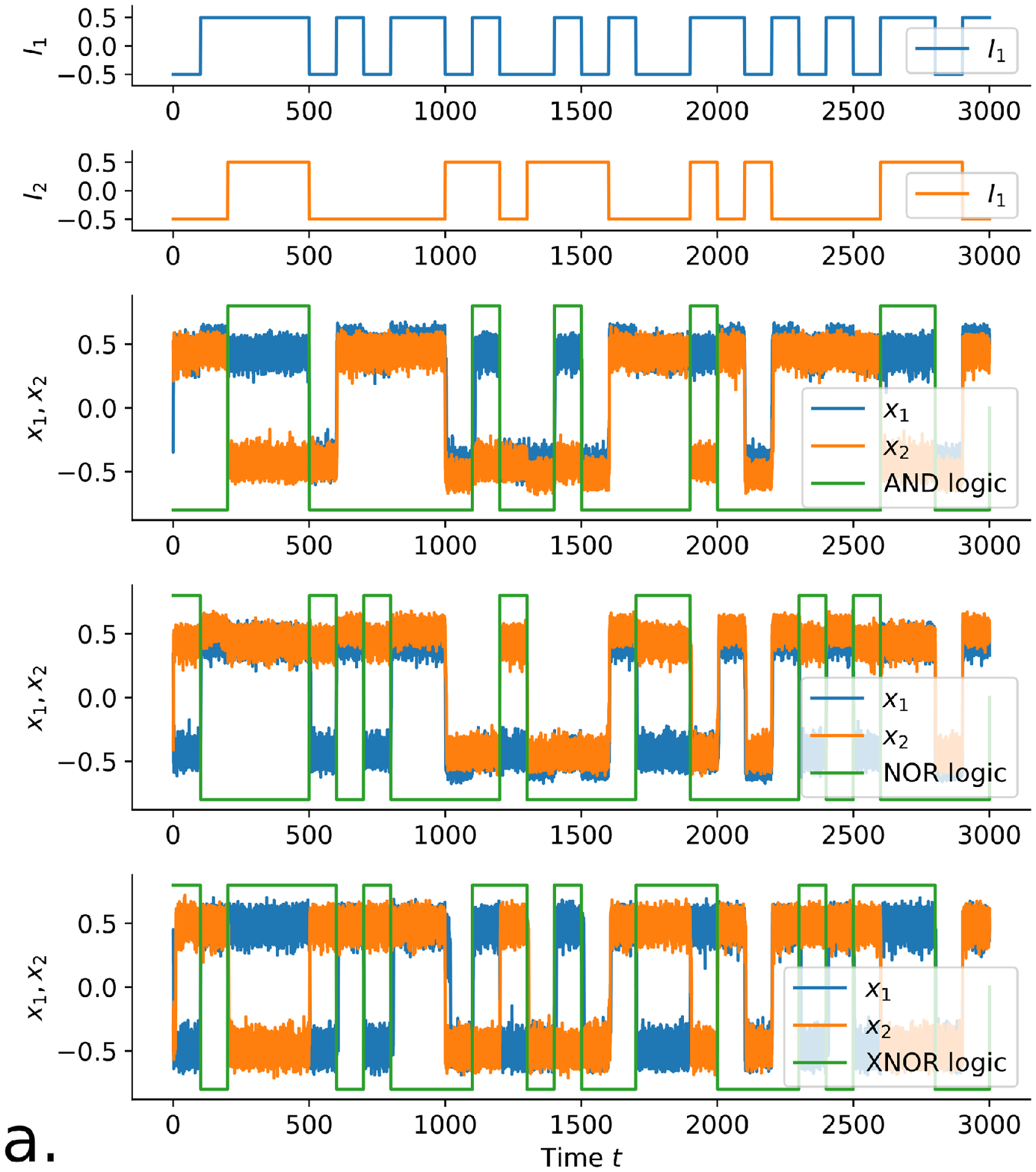}
\includegraphics[width=0.49\linewidth]{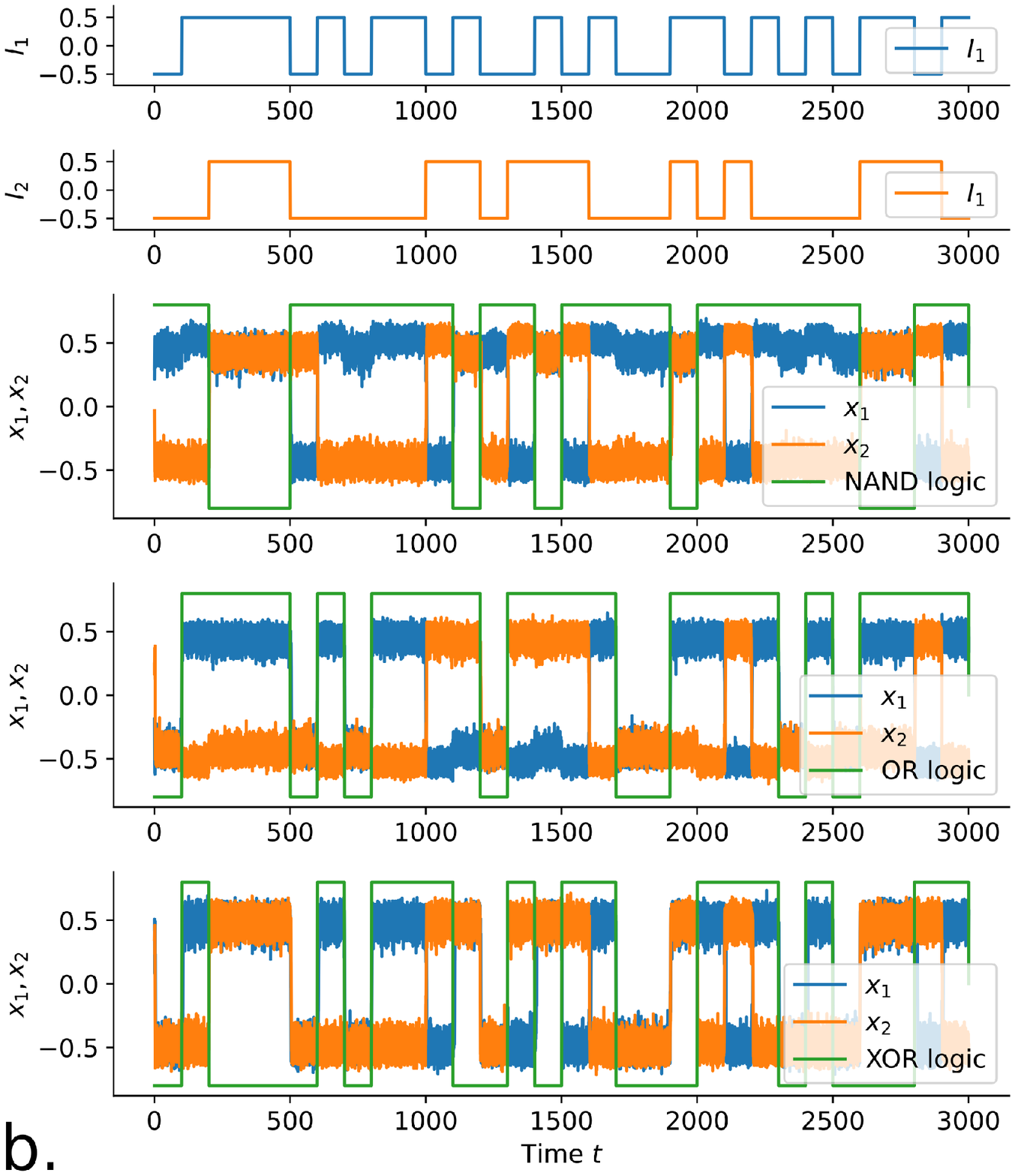}
\vspace{0.2cm}

\caption{ The top two panels show the input streams $I_1$ and $I_2$ that take the value $-0.5$ when the logic input is $0$ and $+0.5$ when the logic input is $1$. The bottom three panels depict the timeseries of the state variables $x_1$ (Blue) and $x_2$ (Orange) obtained from the numerical simulation of (a) the attractive coupling scheme (c.f. Eq.~\ref{eq:att_model}) and (b) the repulsive coupling scheme (c.f. Eq.~\ref{eq:rep_model}). The expected logic output (green) corresponding to each logic operation are plotted over the timeseries as a visual aid. When $x_1$ and $x_2$ are synchronized, the output is interpreted as $0$, when $x_1$ and $x_2$ are anti-synchronized the outputs are interpreted as $1$. Panel 3 shows AND logic operation in the attractive scheme and NAND logic operation in the repulsive scheme for noise strength $D =0.2$, coupling strength $c=1.2$ and bias $b=0.5$. Panel 4 shows NOR logic in the attractive scheme and OR logic in the repulsive scheme for $D=0.2$, $c=1.2$ and $b=-0.5$. Panel 5 shows XNOR operation in the attractive scheme and XOR operation in the repulsive scheme for $D=0.25$, $c=0$ and $b=0$. }
\label{fig:num_trails}
\end{figure*}

\begin{figure*}
\centering
\includegraphics[width=0.43\linewidth]{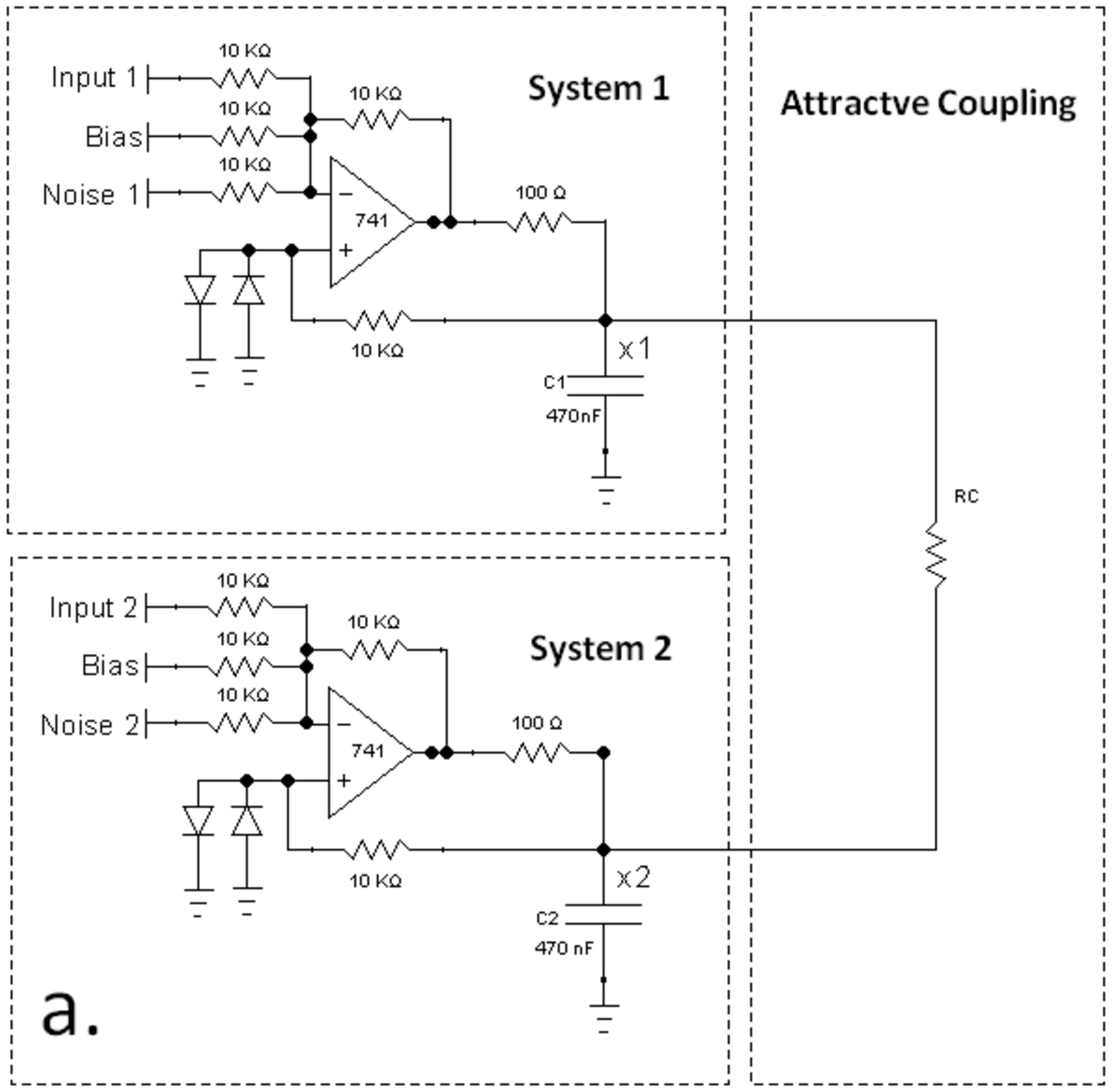} 
\includegraphics[width=0.55\linewidth]{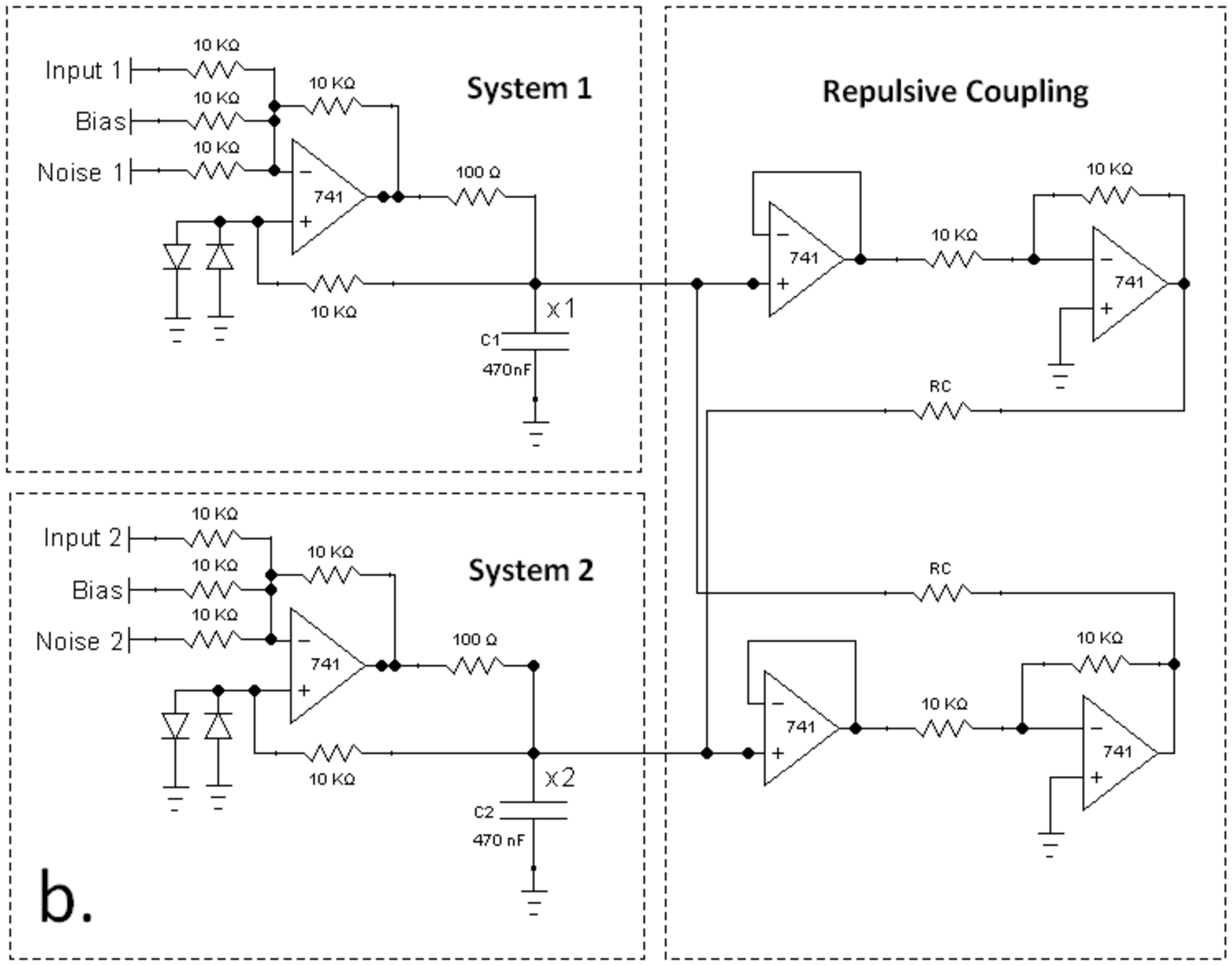}

\caption{The schematic circuit diagram of (a) the attractively coupled system represented by Eqns.~\ref{eq:att_model} and \ref{eq:F_exp}. (b) the repulsively coupled system represented by Eqns.~\ref{eq:rep_model} and \ref{eq:F_exp}. All component values are indicated in the diagram. The diodes used in the circuit are 1N4148 diodes. Both systems are studied for two values of coupling resistances $R_C = 300~\Omega$ and $R= 10~K\Omega$. The system variables $x_1$ and $x_2$ in Eq.~\ref{eq:att_model} and Eq.~\ref{eq:rep_model} are proportional to the voltages $V_1$ and $V_2$ across the capacitors $C_1$ and $C_2$.}
\label{fig:circuit}
\end{figure*}

\begin{figure*}
\centering
\includegraphics[width=0.43\linewidth]{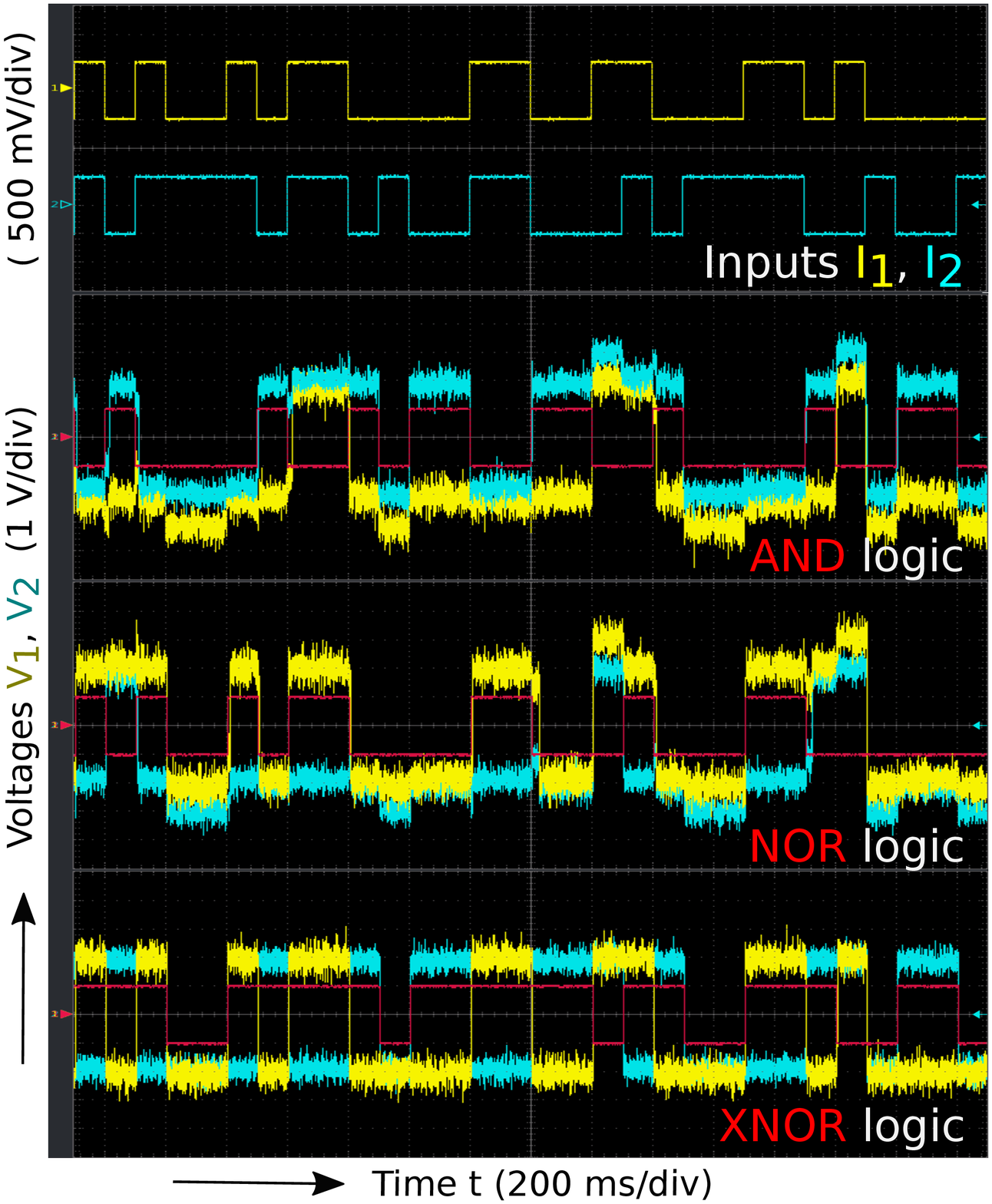}\ \ \hspace{1cm}
\includegraphics[width=0.43\linewidth]{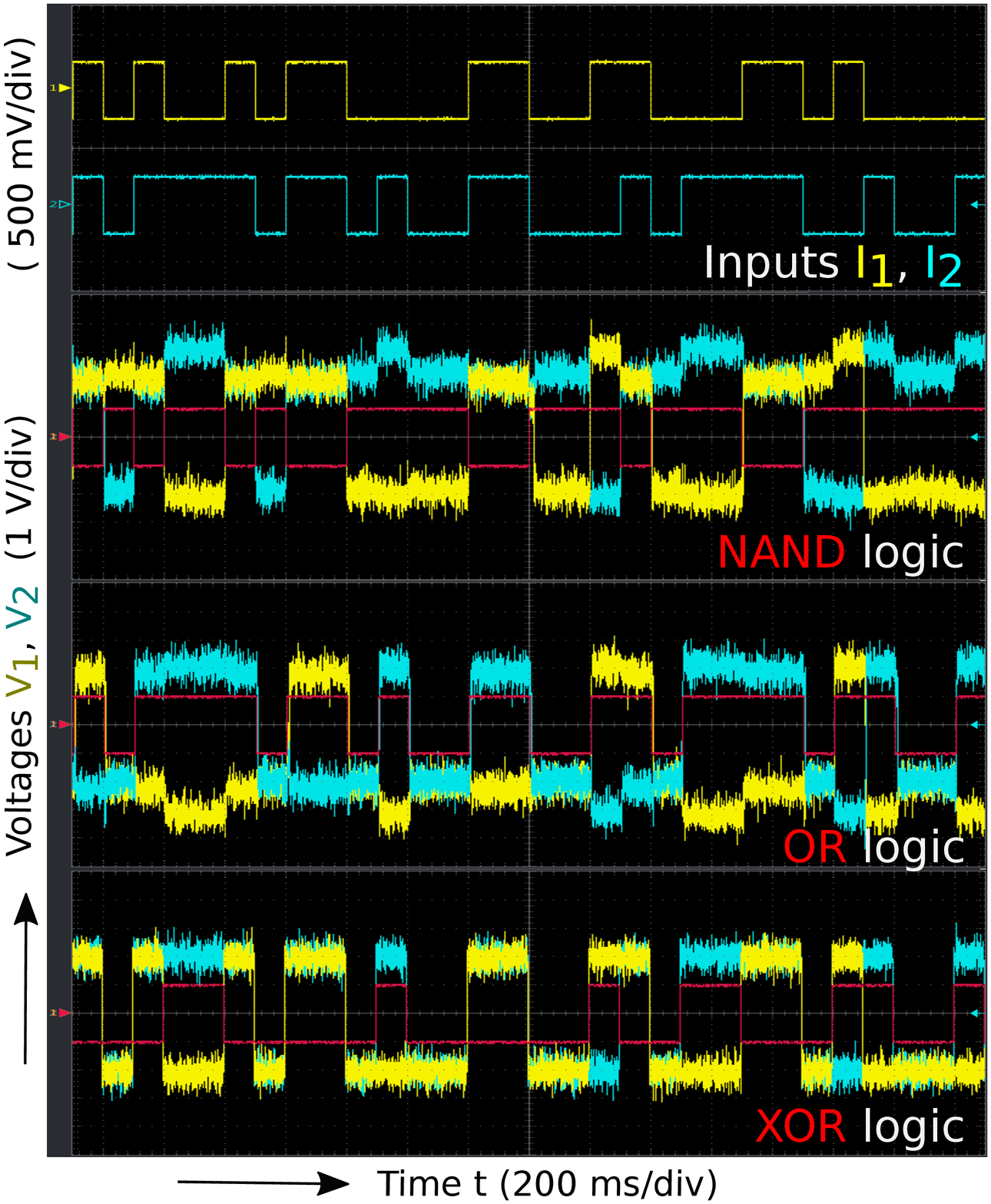}

\caption{Top panels in on both sides are the oscilloscope trace of the logic input signals used to drive the circuits. Left: Oscilloscope traces of the voltages $V_1$ (yellow) and $V_2$ (blue) across capacitors $C_1$ and $C_2$ of the attractive circuit (c.f. Fig.~\ref{fig:circuit}(a)). The expected logical output for each logic gate is presented (in red) as a visual aid. Panel 2 shows AND logic operation obtained for bias $b = 400~mV$, noise strength $D = 0.55~V$ and coupling resistance $R_C = 300~\Omega$. Panel 3 shows NOR logic for $b=-400~mV$, $D = 0.55~V$ and $R_C = 300~\Omega$. In panel 4, XNOR logic is obtained for $b=0~mV$, $D = 0.45~V$ and $R_C = 10~K\Omega$. 
Right: Oscilloscope traces of $V_1$ (yellow) and $V_2$ (blue) from the repulsive circuit (c.f. Fig.~\ref{fig:circuit}(b)). Panel 2 shows NAND logic operation is obtained for bias $b = 400~mV$, noise strength $D = 0.55~V$ and coupling resistance $R_C = 300~\Omega$. Panel 3 shows OR logic is obtained for $b=-400~mV$, $D = 0.55~V$ and $R_C = 300~\Omega$. In panel 4, XOR logic is obtained for $b=0~mV$, $D = 0.45~V$ and $R_C = 10~K\Omega$.}
\label{fig:exp_trails}
\end{figure*}

The scheme is first implemented in silico, by numerically simulating Eqns.~\ref{eq:att_model} and \ref{eq:rep_model} using the Euler-Maruyama method. For the bistable potential, we use the simple cubic function $F(x_i) = 4 (x_i - 5 x_i^3) $. The input signals $I_1$ and $I_2$ are taken to assume the values $-0.5$ for the binary input $0$ and $+0.5$ for binary input $1$. The timetrials of the system variables $x_1$ and $x_2$ thus obtained, are depicted in Fig.~\ref{fig:num_trails} for various values of noise strength $D$, coupling strength $c$ and bias $b$. In this figure, the top two panels, show the input streams $I_1$(blue) and $I_2$(orange) being fed into the two subsystems and the expected logical output(green) of these inputs for each of the six logic operations are overlaid on the timetrails as a visual aid to perceive the correct logical operation. For the six cases depicted, the timetrails of $x_1$ and $x_2$ constantly alternate between the synchronized and anti-synchronized states modulated by the input streams. As defined earlier, the logical output is considered $0$ when the two state variables are in the same potential wells (synchronized) and $1$ when the state variables are in the opposite potential wells (anti-synchronized). From Fig.~\ref{fig:num_trails} it is apparent that robust logic operations are obtained at specific parameter values and coupling forms, for all six types of logic gates.

Next we construct a proof-of-principle electronic implementation of the scheme. Two piecewise-linear bistable circuits were built, using simple passive components and two operational amplifiers(op-amps) as depicted in Fig.~\ref{fig:circuit}. The detailed description and characterization of the bistable circuit used can be found in reference \cite{adomaitiene2008analogue}. The two bistable units were coupled attractively via a resistor as shown in Fig.~\ref{fig:circuit}a or repulsively via inverting amplifiers as shown in Fig.~\ref{fig:circuit}b. The inputs and bias to the bistable system are fed through the inverting input of the op-amps, hence the signals and biases were inverted to stay consistent with the scheme description and the numerical exploration. The non-dimensionalized form of the coupled equation governing the circuits in Fig.~\ref{fig:circuit}a (attractively coupled circuit) and Fig.~\ref{fig:circuit}b (repulsively coupled circuit) assumes the form described in Eq.\ref{eq:att_model} and Eq.\ref{eq:rep_model} respectively. Where, F of bistable unit is given by the piece-wise linear function,

\begin{equation}
    F(x_i) = \left\{
        \begin{array}{ll}
            -(x_i + 1) & \quad x_i < -0.5 \\
            x_i & \quad -0.5 \leq x_i \leq 0.5 \\
	 -(x_i - 1) & \quad x_i > 0.5
        \end{array}
    \right.
    \label{eq:F_exp}
\end{equation}

\noindent
where the state variables $x_1$ and $x_2$ are proportional to the voltages $V_1$ and $V_2$ across the capacitors $C_1$ and $C_2$. The timetrails of the experimental systems were obtained using a Tektronics 2104B Digital Storage Oscilloscope. A high speed data acquisition device (Measurement Computing USB-1616HS) was used to both generate signals ($I_1$, $I_2$, $\eta_1$, $\eta_2$) and collect high throughput voltage data ($V_1$, $V_2$) for further analysis. All signal generation and data collection were done at a rate of $2\times10^4$ samples per second. All specific component values used in the construction of the circuit are indicated in the circuit schematic (c.f. Fig.~\ref{fig:circuit}).

In Fig.~\ref{fig:exp_trails} oscilloscope trails of $V_1$(yellow) and $V_2$(blue) obtained from the attractive circuit (left) and repulsive circuit (right) are presented. The top panel shows the input streams ($I_1$ and $I_2$) driving the coupled system while the remaining six panels clearly show robust logic response akin to the behaviour observed in Fig.~\ref{fig:num_trails}. The expected logic output of the input streams are again overlaid(in red) as a visual aid. The synchronized segments faithfully map to the output $0$ while the anti-synchronized segments map to $1$ for all six logical operations.

The attractive and repulsive coupling schemes yield three logic gates each. The attractive scheme yields AND, NOR and XNOR gates, while the repulsive scheme yields NAND, OR and XOR gate. Notice that the two schemes yield complementary logic operations for the same set of parameter values, suggesting that these two coupling schemes are symmetric counterparts of one another. Within a specific coupling form, the value of the constant bias $b$ (positive, negative or zero) determines the logical operation obtained from the system. The bias ranges and the coupling forms where each of the six logic operations occur, are detailed in Table~\ref{tab:reconf}. 

\begin{table*}
\caption{Logic operations obtained for the various coupling schemes and bias $b$ values.}
\label{tab:reconf} 
\begin{center}
\begin{tabular}{p{4cm}p{2.5cm}p{2.5cm}p{2cm}}
\hline
 & Positive bias & Negative bias & Zero bias\\
\noalign{\smallskip}\hline\noalign{\smallskip}
Attractive coupling & AND & NOR & XNOR\\
Repulsive coupling & NAND & OR & XOR\\
\hline
\end{tabular}
\end{center}
\end{table*}

\begin{figure*}
\centering
\includegraphics[width=0.9\linewidth]{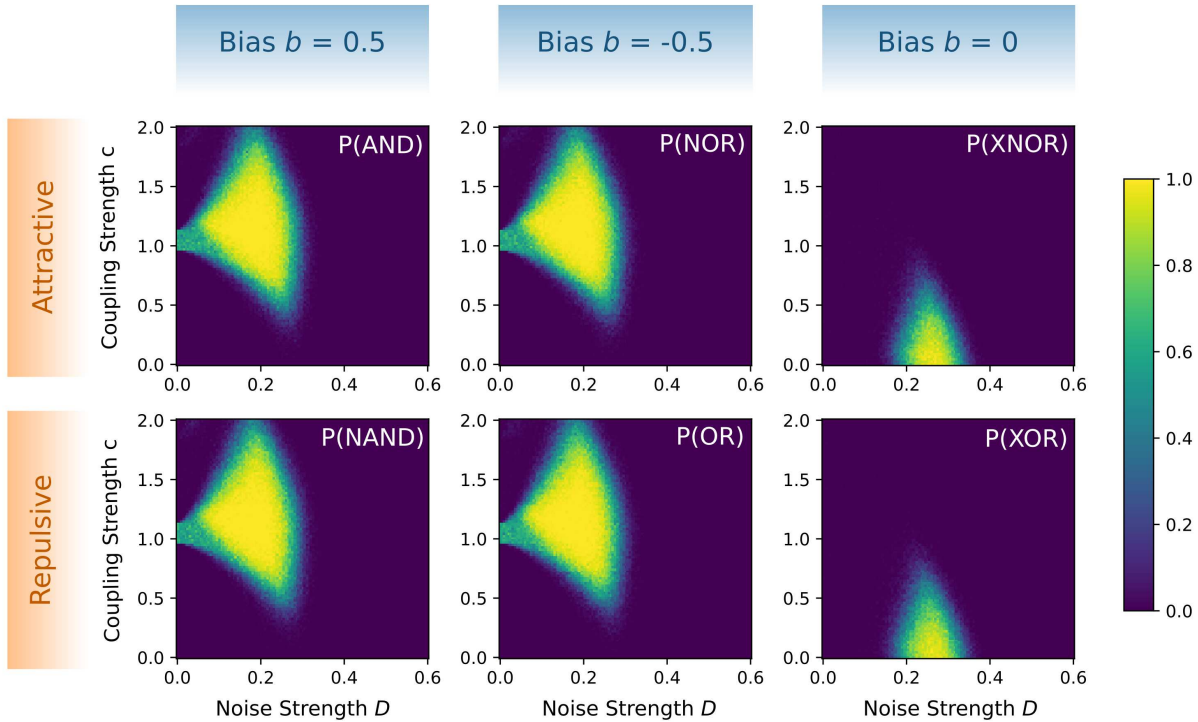}
\caption{Probability of obtaining reliable logic operations $P(logic)$ for all six fundamental logic operations (P(AND), P(NOR), etc.) is plotted as a function of noise strength $D$ and coupling strength $c$. The plots are made for both attractive and repulsive coupling schemes at three specific values of bias $b$, depicting broad regions in parameter space where all six logic operations are consistently obtained.}
\label{fig:num_plogic}
\end{figure*}

\section{Measure of reliability}

Robust operation of the scheme, has been demonstrated both using simulation and experiment for a stream of inputs, with specific values of system parameters. The performance of this system is now quantified with a large number of input sets over a significant section of the parameter space. To do this, the system is subject to a large number of ($I_1 - I_2$) sets, divided into runs where each run consists of a permutation of the four input sets $(0,0)$, $(0,1)$, $(1,0)$, $(1,1)$ and the response of the systems to these inputs are recorded. A run is considered successful, only if the system produced the correct logical output (corresponding to each truth table c.f.~\ref{tab:gates}) throughout the run-time of each signal pulse for all four input sets. The probability of obtaining a specific logic operation $P(logic)$ is then defined as the ratio of successful runs to the total number of runs sampled. A small transient time amounting to one tenth the duration of each input pulse is allowed for the system to respond to each new input set. The $P(logic)$ corresponding to specific logic operations are denoted as P(AND), P(XOR), etc. This measure is then obtained for a large range of parameter values to ascertain the prevalence of reliable logic operations.

In Fig.~\ref{fig:num_plogic}, numerically obtained probabilities of obtaining different logic gates are plotted for a range of noise strengths $D$ and coupling strengths $c$. In these simulations the $P(logic)$ was determined by subjecting the system to 100 runs as described earlier, for each combination of parameter values. Reliable logic operations were seen to occur in large sections of parameter space where the $P(logic)$ value tends to $1$ (bright yellow regions in Fig.~\ref{fig:num_plogic}). The probability plots were made for each of the six logic operations at the corresponding coupling scheme and bias $b$ values mentioned in Table~\ref{tab:reconf}. Thus, we see that \textit{for a window of optimal noise strengths and coupling strengths, all logic operations can be obtained}. The coupling form and bias $b$ act as the control to morph the coupled system from one logic operation to the other. Note that contrary to coupling induced LSR\cite{aravind2018coupling}, where robust logic occurs for all values above a critical coupling strength, we find an optimal range of coupling strengths yield reliable logic operations. 

\begin{figure}
\centering
\includegraphics[width=0.80\linewidth]{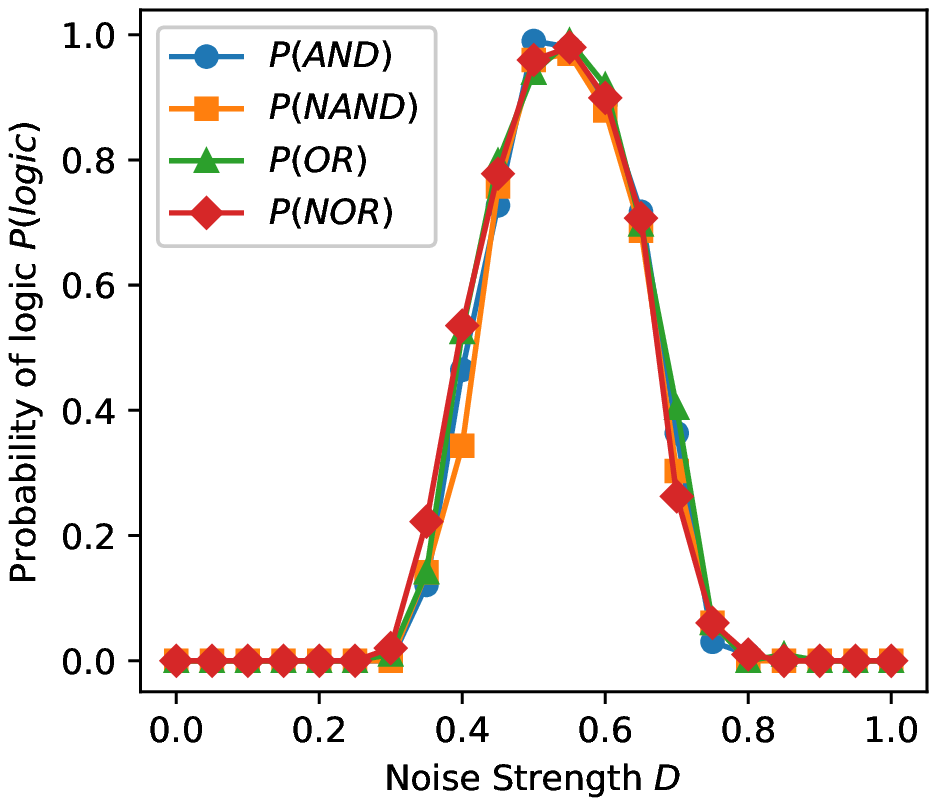}\\
\includegraphics[width=0.80\linewidth]{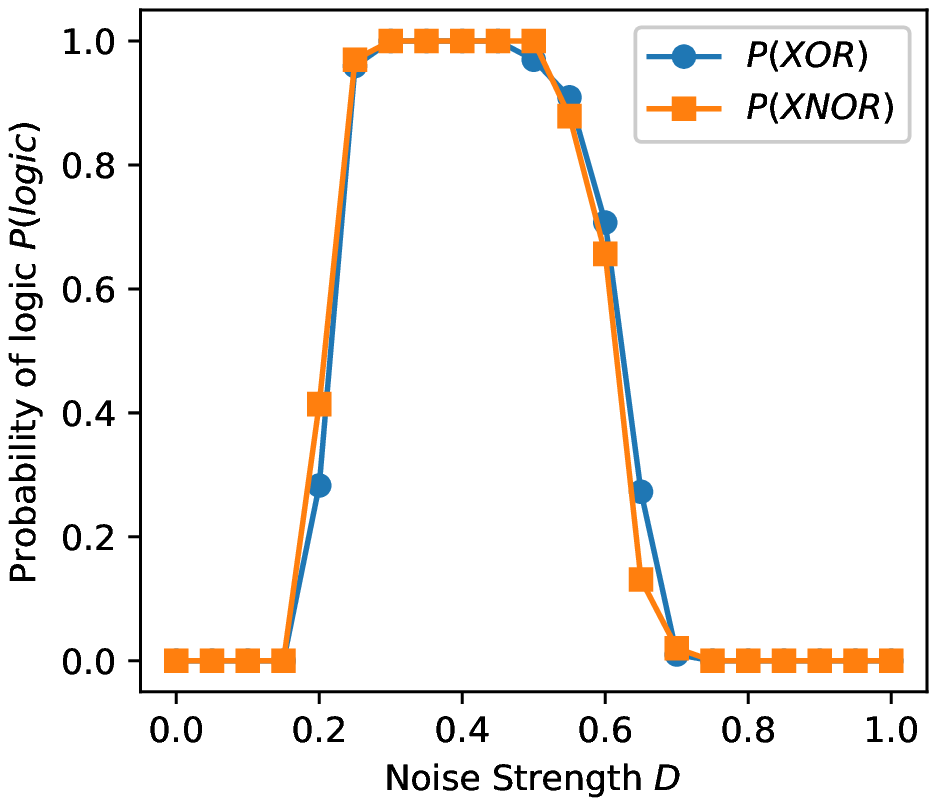}
\caption{Probability of logic P(logic) calculated from experimentally obtained voltage data, plotted as a function of noise strength $D$. Top: For coupling resistance $R_C = 300~\Omega$,  P(AND) and P(NAND) logic plotted for bias $b = 400~mV$. P(NOR) and P(OR) are plotted for $b = -400~mV$. Bottom: For coupling resistance $R_C = 10~K\Omega$,  P(XNOR) and P(XOR) logic plotted for bias $b = 0~mV$.}
\label{fig:exp_plogic}
\end{figure}

To further strengthen this assertion, we also obtain the \textit{same stringent measure} of reliable operations  $P(logic)$, from the experimental implementation of the scheme. This was made possible by interfacing and voltage data analysis through the high speed DAQ. The experimental circuits were driven by an input stream containing 100 runs with a permutation of the four input sets $(0,0)$, $(0,1)$, $(1,0)$, $(1,1)$ in every run. Again, $P(logic)$ is obtained from the ratio of successful runs to total number of runs sampled. The inputs were fed into the system at 10Hz (10 binary inputs per second) and again one tenth of that pulse width was allowed as transient and was not included in the computation of $P(logic)$. Experimentally, two values of coupling resistances were studied, $R=300~\Omega$ and $R=10~K\Omega$. The distribution of $P(logic)$ thus obtained for all the six logic gates are shown in Fig.~\ref{fig:exp_plogic}. A clear maximization of reliable logic operations, occurs for a broad window of noise strength $D$ for all the logic operations. Surprisingly, contrary to the numerical exploration experimentally we find a more robust and broad region of XOR and XNOR operations. Note that the XOR/XNOR operations were the hardest to implement in the past, and their realization necessitated the use of more complicated triple well potentials \cite{xor}, where specific output definitions were assigned for each logic operation. Here, the multiple gates emerge from the collective dynamics of the coupled system and the binary outputs are inferred from their synchronization state.

\section{Conclusion}

A novel scheme to make reconfigurable noise aided logic gates, based on synchronization of coupled nonlinear systems, was introduced. This scheme was implemented both through numerical simulations and electronic experiments. The robustness of the logical operations and the reconfigurability of the scheme was elucidated for both attractive and repulsive coupling. A quantitative measure of performance was used to characterize the region in parameter space of coupling strength and noise strength where reliable logic operations can be obtained. This was done both in simulations and in circuit experiments through live collection and processing of high throughput voltage data. Importantly, all six fundamental logic operations were reliably obtained, using a bias and the coupling form to morph between the logic functionalities. So our results suggest the potential of exploiting synchronization, arising from the interplay of noise and the nature of coupling, to implement flexible logic gates in the presence of a noise floor.

\begin{acknowledgments}
The authors gratefully acknowledge the discussions with Dr. K. Murali, Anna University, Chennai, which helped shape the experimental implementation of the repulsive coupling scheme used in this work. 
\end{acknowledgments}

\end{document}